\documentstyle[a4,12pt]{article}

\newcommand{\ob}[2]{Z_{#1}\times Z_{#2}}
\begin{document}
\makeatletter
\def\siml{\mathrel{\mathpalette\gl@align<}}
\def\simg{\mathrel{\mathpalette\gl@align>}}
\def\gl@align#1#2{\lower.6ex\vbox{\baselineskip\z@skip\lineskip\z@
 \ialign{$\m@th#1\hfill##\hfil$\crcr#2\crcr{\sim}\crcr}}}
\makeatother
\hbadness=10000
\hbadness=10000
\begin{titlepage}
\nopagebreak
\def\thefootnote{\fnsymbol{footnote}}
\begin{flushright}

        {\normalsize
 Kanazawa-94-16\\
August, 1994   }\\
\end{flushright}
\vfill
\begin{center}
\renewcommand{\thefootnote}{\fnsymbol{footnote}}
{\large \bf Non-universal Soft Scalar Masses \\
in Superstring Theories}

\vfill
\vspace{1.1cm}

{\bf Tatsuo Kobayashi,
\footnote[1]{e-mail:kobayasi@hep.s.kanazawa-u.ac.jp}
Daijiro Suematsu,
\footnote[2]{e-mail:suematsu@hep.s.kanazawa-u.ac.jp}
Kiyonori Yamada,
\footnote[3]{e-mail:yamada@hep.s.kanazawa-u.ac.jp}\\
 and \\
 Yoshio Yamagishi
\footnote[4]{e-mail:yamagisi@hep.s.kanazawa-u.ac.jp}
}

\vspace{1.1cm}
       Department of Physics, Kanazawa University, \\

       Kanazawa, 920-11, Japan \\

\vfill
\end{center}
\vspace{1.1cm}

\vfill
\nopagebreak
\begin{abstract}
We study soft scalar masses comparing with gaugino masses in 4-dimensional
string models.
In general non-universal soft masses are derived in orbifold models.
We give conditions on modular weights to lead to the large non-universality
in the soft scalar masses.
This non-universality is applied to the unification of the gauge coupling
constants in the minimal string model.

\end{abstract}

\vfill
\end{titlepage}
\pagestyle{plain}
\newpage
\voffset = 0.5 cm

\vspace{0.8 cm}
\leftline{\large \bf 1. Introduction}
\vspace{0.8 cm}

Supersymmetric models are very interesting as the unified theory.
Local supersymmetry (SUSY) breaking induces soft SUSY breaking terms such as
gaugino masses, scalar masses and trilinear (A-terms) and bilinear (B-terms)
couplings of scalar fields in global SUSY models \cite{Nilles}.
The values of these soft terms determine the phenomenological properties of
the models.

Superstring theories are only known candidates for the unified theory, when
we take gravity into account.
By now based on 4-dim string models like Calabi-Yau models \cite{CY},
orbifold models [3--6] and so on, we know a K\"ahler potential and a gauge
kinetic function of supergravity as their effective theories.
In recent papers [7--10] the soft SUSY breaking terms derived from
superstring theories are studied assuming that F-terms of a dilaton field
$S$ and moduli fields contribute to the SUSY breaking.
There the soft scalar masses are non-universal at the string scale
$M_{\rm st}=3.73 \times 10^{17}$GeV \cite{Kaplunovsky}, although one usually
assumes the universality of all the soft terms in SUSY models.
This non-universality affects phenomenological features \cite{KSY,Non}.
For example, in ref. \cite{KSY} it is shown that a unification scale of the
SU(3) and SU(2) gauge couplings in the minimal supersymmetric standard model
(MSSM) is sensitive to the non-universality such that orders of
magnitudes of the soft masses differ one another.
In general the non-universality increases the unification scale.
This seems desirable for the superstring unification.

However, the orders of the soft scalar masses derived in refs.
\cite{Carlos,Brignole} seem not to be so different each other.
In ref.\cite{Brignole} one assumes that only the dilation field and an
overall modulus field $T$ contribute to the SUSY breaking and parametrizes
an unknown goldstino field by a goldstino angle $\theta$.
Generally the orbifold models have three independent moduli fields.
In this paper we study the case where the three moduli fields as well as $S$
contribute to the SUSY breaking in the orbifold models, and investigate the
possibility to obtain the soft scalar masses whose orders of magnitudes are
different one another.
We apply such a non-universality to the unification of the gauge couplings
in a minimal string model \cite{Ibanez,MSM,MSM2,MSM3} as an interesting
example.
The minimal string model is a string vacuum which has the same massless
spectrum as the MSSM.

This paper is organized as follows.
In section two we review the soft SUSY breaking terms obtained in ref.
\cite{KL} and also their parameterization following ref.\cite{Brignole}.
In section three we reformulate the soft terms taking account of the three
moduli fields so as to obtain the scalar masses whose orders of magnitudes
are different one another.
Conditions to yield such hierarchical soft masses are given.
In section four such a non-universality is applied to the minimal string
unification.
Using threshold corrections of string massive modes
\cite{Thres,Derendinger}, we investigate whether orbifold models can
realize the observed low energy values of gauge coupling within the
framework of the MSSM.
Section five is devoted to conclusions and discussions.

\vspace{0.8 cm}
\leftline{\large \bf 2. Soft masses}
\vspace{0.8 cm}

In this section, we review the derivation of soft SUSY breaking terms
from the superstring models following ref.\cite{KL}.
Here we assume that the SUSY breaking occurs in terms of only the F-terms
of the dilaton field and the moduli fields $T^i$ $(i=1 \sim 3)$, where $T^i$
corresponds to the $i$-th one among three moduli of the 6-dim orbifolds [3-6].
We represent $S$ and $T^i$ by $\Phi^m$ ($m=0\sim3$), where $\Phi^0$ is $S$
and $\Phi^1=T^1$, etc.
Supergravity theories are determined in terms of
the K\"ahler potential $K$, the superpotential $W$ and the gauge kinetic
function $f_a$.
The K\"ahler potential and the superpotential are expressed as follows,
$$K=\kappa^{-2}\hat K(\Phi,\bar \Phi)+
K(\Phi,\bar \Phi)_{I{\bar J}}Q^I\bar Q^{\bar J}
+({1 \over 2}H(\Phi,\bar \Phi)_{IJ}Q^IQ^J+{\rm h.c.})+\cdots,
\eqno(2.1a)$$
$$W=\hat W(\Phi)+{1 \over 2}\tilde \mu(\Phi)_{IJ}Q^IQ^J+\cdots,
\eqno(2.1b)$$
where $\kappa^2=8\pi/M_{\rm pl}^2$ and $Q^I$ are chiral superfields.
The ellipses stand for terms of higher orders in $Q^I$.
Using these, we can write down the scalar potential $V$ as follows,
$$V=\kappa^{-2}e^G[G_\alpha(G^{-1})^{\alpha \bar \beta}G_{\bar \beta}
-3\kappa^{-2}]+({\rm D-term}),
\eqno(2.2)$$
where $G=K+\kappa^{-2}\log \kappa^6 |W|^2$ and the indices $\alpha$ and
$\beta$ denote $Q^I$ as well as $\Phi^m$.
Here we restrict ourselves to the case without D-term contribution.
The gravitino mass $m_{3/2}$ is written as
$$m_{3/2}=\kappa^2e^{\hat K}|\hat W|.
\eqno(2.3)$$
In (2.2) we take the flat limit $M_{\rm pl}\rightarrow \infty$ while
$m_{3/2}$ is fixed, so that the soft scalar masses $m_{IJ}$ are derived as
$$m_{I{\bar J}}^2=m_{3/2}^2K_{I{\bar J}}
-F^m{\bar F}^{\bar n}[\partial_m\partial_{\bar n}K_{I{\bar J}}
-(\partial_{\bar n}K_{K{\bar J}})K^{K{\bar L}}(\partial_mK_{I{\bar L}})]
+\kappa^2V_0K_{I{\bar J}},
(2.4)$$
where $F^m$ are F-terms of $\Phi^m$, $\partial_m$ denote
$\partial/\partial \Phi^m$ and $V_0$ is the cosmological constant
expressed as
$$V_0=\kappa^{-2}(F^m{\bar F}^{\bar n} \partial_m \partial_{\bar n}
\hat K-3m_{3/2}^2).
\eqno(2.5)$$
Hereafter the gravitational coupling $\kappa$ is set to one.
The gaugino masses $M_a$ are derived through the following
equation,
$$M_a={1 \over 2}{\bar F}^{\bar m}\partial_{\bar m}\log Re f_a,
\eqno(2.6)$$
where the subscript $a$ represents a gauge group.

The orbifold models give the K\"ahler potential at the one-loop level
as follows \cite{Derendinger},
$$K=-\log Y-\sum_i \log (T^i+\bar T^i)
+\prod_i (T^i+\bar T^i)^{n^i_I}Q^I{\bar Q}^{\bar I},$$
$$ Y=S+\bar S-\sum_i{\delta^i_{GS}\over 8\pi^2}\log(T^i+\bar T^i),
\eqno(2.7)$$
where $\delta^i_{GS}$ are coefficients of the Green-Schwarz mechanism to
cancel duality anomalies \cite{GS,Derendinger2,Derendinger} and $n^i_I$ are
modular weights corresponding to $Q^I$ \cite{Dixon,Ibanez}.
The untwisted sector associated with the $j$-th plane has
$n^i_I=-\delta^i_j$.
The twisted sector with a twist $v^i$
$\displaystyle (0 \leq v^i \leq 1, \sum^3_{i=1}v^i=1)$
has $n^i_I=v^i-1$ if $v^i \neq 0$,
and $n^i_I=0$ if $v^i=0$.
For the notation of the orbifolds and their twists, we follow
refs.\cite{KO3,ZNM} except some permutation of elements $v^i$.
The addition of oscillators change the modular weights by one.
We have also the following gauge kinetic function:
$$f_a=k_aS-{1 \over 16\pi^2}\sum_i (b'^i_a-k_a\delta^i_{GS})\log
|\eta(T^i)|^4,
\eqno(2.8)$$
where the second term is a threshold correction due to string massive modes
\cite{Thres,Derendinger}.
Here $b'^i_a$ is a duality anomaly coefficient and $\eta(T)$ is a Dedekind
function.
In addition $k_a$ denotes a level of each gauge group.
In general the 4-dim string models lead to $k_a=1$
for non-abelian gauge groups.

In ref.\cite{Brignole} the case of the overall modulus $T=T^i$ is discussed.
This case is characterized by sums as $\displaystyle n=\sum_i n^i,
\ b'_a=\sum_ib'^i_a$
and $\displaystyle \delta_{GS}=\sum_i \delta_{GS}^i$.
Then we have the cosmological constant as,
$$V_0=-3m_{3/2}^2+{1 \over Y^2}|F^0-
{\delta_{GS} \over 8\pi^2(T+\bar T)}F^T|^2+
{3 \over (T+\bar T)^2}(1-{\delta_{GS} \over 24 \pi^2Y})|F^T|^2,
\eqno(2.9)$$
where $F^T$ is the F-term corresponding to the overall modulus $T$.
If $V_0=0$, one can parametrize the unknown F-terms by the goldstino angle
$\theta$ as follows,
$${1 \over Y}(F^0-{\delta_{GS} \over 8\pi^2(T+\bar T)}F^T)=
\sqrt 3 m_{3/2}\sin \theta,
\eqno(2.10a)$$
$${\sqrt {3 (1-a)}\over T+\bar T}F^T=\sqrt 3 m_{3/2} \cos \theta,
\eqno(2.10b)$$
where $a=\delta_{GS} /24 \pi^2Y$.
The masses of the scalar superpartners and the gauginos are expressed as,
$$m^2=m_{3/2}^2(1+{n \over 1-a}\cos^2 \theta),
\eqno(2.11a)$$
$$M_a={\sqrt 3 \over f_a}m_{3/2}(k_a Re S \sin \theta +\cos \theta
{b'_a-k_a\delta_{GS} \over 32 \pi^3 \sqrt {3(1-a)}}(T+\bar T)
\hat G_2(T,\bar T)),
\eqno(2.11b)$$
where $\hat G_2(T,\bar T)$ is the Eisenstein function;
$\hat G_2(T,\bar T)=-4\eta(T)^{-1}(\partial \eta(T) /\partial T)
-2\pi/(T+\bar T)$\footnote{Several kinds of modular functions are shown
in ref.\cite{Cvetic}}.
The non-oscillated sectors are allowed to have $n=-1$ and $-2$.
Thus the different modular weights lead to the non-universal soft masses in
(2.11a).
However most of these masses are of $O(m_{3/2})$.
Also the gaugino masses are the same order as $m_{3/2}$ unless
$\sin \theta=0$.
Even if the orders of the scalar masses could differ one another at
$M_{\rm st}$, loop effects due to the gauginos dilute the
non-universality at $M_Z$ so as to result in the scalar masses with the
same order, i.e., $O(m_{3/2})$.
Therefore we are interested in the case with $\sin \theta =0$ in order to
produce the fair non-universality in the soft scalar masses.
In this case, unfortunately matter fields with $n \leq -2$ are not allowed,
because these modular weights lead to the imaginary masses in (2.11a).
When we restrict ourselves to the non-oscillated states, the soft scalar
masses are completely universal and of order of $\sqrt {|a|}m_{3/2}$.
Note that $a$ should be negative so that this mass is well defined.
As a result the soft scalar masses are the same order in any case.
That might become a very severe constraint on the SUSY models inspired by
the superstring theories.

Oscillated states with $n=0$ seems to yield the non-universality in the
mass spectrum of the scalar fields.
However, the presence of such states is restricted in some cases as shown
in refs.\cite{MSM2,MSM3}.
Moreover Yukawa couplings of such states are often forbidden as
renormalisable couplings.
Therefore we study another possibility to cause the fairly non-universal
soft scalar masses in the following section.

\vspace{0.8 cm}
\leftline{\large \bf 3. Soft masses in case of three moduli}
\vspace{0.8 cm}

We study more general case than the previous section by taking into account
three independent moduli fields $T^i$ instead of the overall moduli.
In this case the cosmological constant $V_0$ is written as,
$$V_0=-3m_{3/2}+{1 \over Y^2}|F^0-\sum^3_{i=1}
{\delta_{GS}^i \over 8 \pi^2(T^i+\bar T^i)}
F^i|^2+\sum^3_{i=1} {1-a_i \over (T^i+\bar T^i)^2}|F^i|^2,
\eqno(3.1)$$
where $a_i=\delta_{GS}^i/8\pi^2Y$ and it is estimated as $|a_i|<<1$.
Here we parametrize the unknown F-terms as follows,
$${ 1 \over Y}(F^0-\sum_i{\delta_{GS}^i \over 8 \pi^2(T^i+\bar T^i)}F^i) =
\sqrt 3 m_{3/2} \sin \theta,
\eqno(3.2a)$$
$${\sqrt{1-a_i} \over T^i + \bar T^i}F^i=\sqrt 3 m_{3/2} \cos \theta
\Theta_i,
\eqno(3.2b)$$
where $\Theta_1=\sin \theta' \sin \theta''$,
$\Theta_2=\sin \theta' \cos \theta''$ and $\Theta_3=\cos \theta'$.
We replace $m_{3/2}$ in (3.2) by $Cm_{3/2}$, unless $V_0=0$.
In ref.\cite{Choi} it is indicated that $V_0$ should be negative taking
account of loop effects of observable sectors.
If $V_0$ is negative, $C$ is less than one.
We can write down the soft masses as,
$$m^2=m_{3/2}^2(1+C^2\cos^2 \theta \sum^3_{i=1} {3n^i \over 1-a_i}
\Theta_i^2) +2m_{3/2}^2(C^2-1),
\eqno(3.3a)$$
$$M_a={\sqrt 3 \over f_a}Cm_{3/2}[k_aRe S \sin \theta +
\cos \theta \sum^3_{i=1}{b'^i_a-k_a\delta_{GS}^i \over \sqrt {1-a_i}}
D_i(T^i,\bar T^i)\Theta_i],
\eqno(3.3b)$$
where $D_i(T^i,\bar T^i)=(T^i+\bar T^i)\hat G_2(T^i,\bar T^i)/32\pi^3$.
It is remarkable that the summation in (3.3b) is taken on the moduli
contributing to the threshold corrections.

We are interested in the case with $m>M_a$, otherwise the non-universality
of the soft scalar masses is diluted by the loop effects from $M_a$.
To realize such cases we investigate the modular weights which guarantee
the real soft scalar masses when $\sin \theta=0$ and $a_i<0$.
The first column of Table 1 (2) shows the modular weights of the $Z_N$
($Z_N \times Z_M$) orbifold models without oscillators in the case where
$C=1$.
Only the modular weights corresponding to the matter sectors
\cite{KO,KO3,KO2} are shown in
the tables and the modular weights for the anti-matter sectors are omitted.
The underlines represent any permutation of the elements.
The first three rows in Table 1 correspond to the untwisted sectors, which
are omitted in Table 2, and the others correspond to the twisted sectors.
For the twisted sectors all the modular weights with
$\displaystyle \sum_i n_i=-1$ are
allowed under certain conditions as shown in the tables.
It is remarkable that the modular weights $-(\underline {5,5,2})/6$ are
forgiven under certain conditions, although
the case of the overall modulus forbids all the modular weights with
$\displaystyle \sum_i n_i=-2$.
For the $Z_3$ and $Z_7$ orbifold models, any twisted matter fields is not
allowed.
The $Z_4$ and $Z_8$-I orbifolds have only the modular weight $-(1,1,0)/2$
in the twisted sectors.
Further, all the modular weights are not forgiven simultaneously.
Actually both modular weights of $(0,0,-1)$ and $-(5,5,2)/6$ cannot
guarantee the reality of scalar masses.
For the $Z_N$ orbifold models all the modular weights with $n=-1$ are
allowed simultaneously in the case where $\sin^2 \theta'=2/3$ and
$\sin^2 \theta''=1/2$.

We can easily obtain the conditions in the case where $C\neq 1$.
For example, the modular weight $(-1,0,0)$ can derive the real scalar
masses under the following condition,
$$\sin^2\theta' \sin^2 \theta'' \leq {2 \over 3}-{1 \over 3C^2}.
\eqno(3.4)$$
Thus the modular weight $(-1,0,0)$ is impossible in the case where
$C^2<1/2$ or $V_0<-3m_{3/2}^2/2$.
If $V_0$ is negative, i.e., $C<1$, the modular weight $-(5,5,2)/6$ is not
allowed and all the modular weights with $n=-1$ cannot satisfy the
conditions for the real scalar masses simultaneously.

Here we study in more detail the soft scalar masses in the case where
$\sin \theta=0$ and $C=1$.
Using (3.3a) we find the representative orders of the scalar masses as
$O(m_{3/2})$ and $O(\sqrt {|a_i|}m_{3/2})$.
We have $a_i \approx 5.0 \times \delta_{GS}^i \times 10^{-3}$ when we
use the unified gauge coupling as $\alpha_X^{-1} \approx 25$.
As an example, we consider the scalar fields with modular weights
($-1,0,0$), $(0,-1,0)$ and $-(1,1,0)/2$.
Most of the orbifold models have these modular weights.
Then we take the angles as $\sin \theta=0$, $\sin^2 \theta'=1/3$ and
$\cos^2 \theta''=1$.
The scalar fields with $n_i=-(1,1,0)/2$ and $(-1,0,0)$ have the soft masses
as $m^2=m_{3/2}^2/2$ and $m_{3/2}^2$, respectively.
On the other hand, the mass corresponding to $n_i=(0,-1,0)$ is obtained as
$m^2=a_2m_{3/2}^2$.
Thus we can derive the different orders of the non-universality among the
soft scalar masses in the case where we take into account the three moduli
fields.
Similarly we can also obtain the fair non-universality under certain
angles $\theta'$ and $\theta''$ for other combinations of the modular
weights in the case with some values of $C$.
The largest mass is of order of $m_{3/2}$.
Such a situation is impossible in the case of the
overall modulus with $\sin \theta=0$.

It is notable that the mass of order of $m_{3/2}$ is not obtained
when some combinations of the modular weights constrain the angles
$\theta'$ and $\theta''$ severely.
For example, we take the case where the matter fields with $n_i=(0,0,-1)$
are included in addition to the above combination of the modular weights.
This case is only possible under the condition where $\sin^2 \theta'=2/3$
and $\sin^2 \theta''=1/2$ as said before.
These angles result in the scalar masses of order of
$O(\sqrt {|a_i|} m_{3/2})$ at most.
Similarly the scalar mass of $n_i=-(5,5,2)/6$ is less than
$O(\sqrt {|a_i|} m_{3/2})$, because the matter fields is allowed at
$\sin \theta'=0$.
Thus we cannot derive the soft scalar masses of order of $O(m_{3/2})$
in the case where the angles $\theta'$ and $\theta''$ are constrained
severely.

Next we estimate the gaugino masses, which should be small enough not to
dilute the non-universality of the soft scalar masses by loop effects.
In (3.3b) the term $D_i(T^i,\bar T^i)$ takes values as
$D_i(T^i,\bar T^i)=1.5\times 10^{-3}$, $2.7 \times 10^{-3}$,
$6.0 \times 10^{-2}$ and $6.6\times 10^{-1} $under $T^i=1.2$, $5.0$, $10$
and 100, respectively.
In (3.3b) the first term proportional to $\sin \theta$ contributes mainly
to the gaugino masses in the case where $\sin \theta >O(10^{-3})$ and
$T\sim O(1)$.
This is because $\sqrt 3/f_a\approx 1.4$ for $k_a=1$ and
$\alpha_X^{-1} \approx 25$ and then $k_a Re S \sim f_a \sim O(1)$.
The condition $\sin \theta <O(1/10)$ should be satisfied in order not to
dilute the above mentioned non-universality of the soft scalar masses.
The large value of $T^i$ like $T_i>O(100)$ seems undesirable for the
non-universality of the soft scalar masses.

\vspace{0.8 cm}
\leftline{\large \bf 4. Minimal string unification}
\vspace{0.8 cm}

In ref.\cite{KSY} it is shown that the unification scale $M_X$ of the
SU(2) and SU(3) gauge coupling constants is sensitive to the
non-universality of the soft masses in the MSSM.
In that paper $M_X$ is estimated using the measured gauge coupling
constants at $M_Z$.
The unification scale $M_X$ increases in the most cases of the
non-universality.
Especially the highest $M_X$ is realized in the case where all the doublet
scalar fields under SU(2) are heavier than the singlet ones.
This type of the non-universality corresponds to Case III in
ref.\cite{KSY}.
In this section we apply the non-universality discussed above to the
minimal string unification as a typical interesting example.
Here we concentrate ourselves on the non-universal case where all the
doublet scalar fields are heavier than singlet ones.
Note that the gauge coupling of U(1)$_Y$ $\alpha_1^{-1}$ is not always
unified at $M_X$ with the other couplings, because the string theories
can predict not only $k_1=5/3$ but also other value.

The running gauge coupling constants $\alpha_a^{-1}$ of the MSSM at $\mu$
are expressed as follows \cite{Thres,Derendinger},
$$\alpha_a^{-1}(\mu)=\alpha_{\rm st}^{-1} +{b_a \over 4 \pi k_a}
\log{M_{\rm st}^2 \over \mu^2}-\sum_i
{b'^i_a-k_a \delta_{GS}^i \over 4 \pi k_a}\log [(T^i+\bar T^i)|\eta(T^i)|^4]
,\eqno(4.1)$$
where $\alpha_{\rm st}^{-1}$ is the universal string coupling at
$M_{\rm st}$ and $b_a$ is the one-loop $\beta$-fuction coefficient of the
MSSM, i.e., $b_3=-3$, $b_2=1$ and $b_1=11$.
The last term in (4.1) represents the threshold correction due to the
string massive modes and the duality anomaly coefficient $b'^i_a$ is
written as
$$b'^i_a=-C(G_a)+\sum T(\underline{R_a})(1+2n^i),
\eqno(4.2)$$
where $C(G_a)$ is a quadratic Casimir of an adjoint representation and
$T(\underline{R_a})$ is a Dynkin index of an $\underline{R_a}$
representation, i.e.,
$T(\underline {R_a})=C(\underline{R_a})\dim(\underline {R_a})
/\dim (G_a)$.
Using eq. (4.1) we can derive the relation between $M_X$ and $M_{\rm st}$
as \cite{Ibanez},
$$8\log {M_X \over M_{\rm st}}=\sum_i(b'^i_3-b'^i_2)\log [(T^i+\bar T^i)
|\eta(T^i)|^4].
\eqno(4.3)$$
It is remarkable that $\log [(T^i+\bar T^i)|\eta(T^i)|^4]$ is negative for
any value of $T^i$.
The unification scale $M_X$ is always less than $M_{\rm st}$ under the
condition that the soft masses is less than 10TeV even in the non-universal
case \cite{KSY}.

The $Z_3$ and $Z_7$ orbifold models do not have the $T$-dependent threshold
corrections and so these orbifolds cannot yield the minimal string models
consistent with the experiments.
For the other $Z_N$ orbifold models except $Z_6$-II, only the third modulus
$T^3$ contributes to the threshold correction.
In these orbifold models, the duality anomaly coefficients are required
to satisfy $b'^3_3>b'^3_2$ in order to result in $M_X<M_{\rm st}$.
First of all we consider the case where the matter fields have the modular
weights $-(1,1,0)/2$, $(-1,0,0)$ and $(0,-1,0)$ as discussed in the
previous section.
The third elements of these modular weights vanish and we obtain
$b'^3_3-b'^3_2=-2$.
In this case we cannot realize the measured gauge couplings at $M_Z$.

In order to avoid such a situation, we need the modular weights with the
non-vanishing third element, which are $(0,0,-1)$ and $-(5,5,2)/6$.
The former belongs to the untwisted sector and the latter exists
only in the $Z_6$-I and $Z_{12}$-I orbifold models.
If the SU(2) doublet scalar fields have such modular weights,
the difference $b'^3_3-b'^3_2$ increases.
Now we are considering the case where the doublet fields are heavier than
the singlet fields.
Thus it seems desirable that the scalar fields masses associated with
$n_i=(0,0,-1)$ or $-(5,5,2)/6$ correspond to the doublets and their masses
are of order of $O(m_{3/2})$.
However the modular weight $-(5,5,2)/6$ cannot derive the soft scalar masses
 of order $O(m_{3/2})$ and is not desirable for the above scenario.
Similarly the simultaneous presence of the modular weights $(0,0,-1)$ and
$-(1,1,0)/2$ forbids the scalar fields with $(0,0,-1)$ to have the masses
 of order $O(m_{3/2})$, because the angle $\theta'$ is constrained as
$\sin^2 \theta'=2/3$.
Therefore we cannot realize the minimal string models with the non-universal
soft scalar masses of Case III using the twisted sectors of the $Z_4$ and
$Z_8$-I orbifold models, where only $n=-(1,1,0)/2$ is allowed among the
twisted sectors.
The $Z_6$-I orbifold models are not promising, either.
Although we can use only the untwisted sectors, it seems unrealistic that
the massless spectrum consists of only the untwisted sectors.
The scalar fields with $n=(0,0,-1)$ can obtain the masses of order
$O(m_{3/2})$ with the presence of some twisted matter fields in the $Z_8$-II
 and $Z_{12}$-I, II orbifold models, because these orbifolds forgive a
several types of the modular weights.

Next, we consider the $Z_6$-II orbifold models, where the $T^2$ and $T^3$
contribute to the threshold corrections.
They have several types of modular weights which have non-vanishing
elements on the second and third.
For example we take the modular weights $-(1,1,0)/2$ and $-(2,0,4)/6$, and
assign the former to the doublet fields and the latter to the singlet
fields.
We assume that $n_i=(-1,-1,0)/2$ derives the scalar mass
$\sqrt 3 m_{3/2}/2$ and the other scalar mass vanish in the case with
$\sin^2 \theta'=1/2$ and $\sin^2 \theta''=0$.
The latter scalar fields gain the mass of order of the gaugino mass
at $M_Z$ by loop effects.
If $T^2=T^3$, we obtain $b'^2_3+b'^3_3-b'^2_2-b'^3_2=-2$ under the above
assignment of the modular weights to the matter fields.
Therefore we cannot have the gauge couplings consistent with the
experiments.
Then we consider the case where only $T^2$ contributes to the threshold
correction, i.e., $T^2>T^3$.
In this case we obtain $b'^2_3-b'^2_2=2$.
The results of ref.\cite{KSY} shows that the unification scale of Case III
is estimated as
$\log_{10}M_X({\rm GeV})=17.0$, 17.1, 17.2 and 17.3 in the case where
the doublet superpartners have the mass of 1.3, 2.0, 3.2 and 5.0 TeV,
respectively, while the gauginos and the singlet superpartners have the
masses of 100GeV.
In this case the doublet scalar masses correspond to
$\sqrt 3 m_{3/2}/2$ and then we can easily estimate $m_{3/2}$.
Using (4.3) we obtain the desirable values of $T^2$ as $T^2=7.5$, 6.5, 5.5
and 4.5 in the case with $\log_{10}M_X({\rm GeV})=17.0$, 17.1, 17.2
and 17.3, respectively.
Further these values of $T^2$ derive $D_2(T^2,\bar T^2)=0.043$, 0.037,
0.030 and 0.024, respectively.
In the case with $\sin \theta < 10^{-2}$, we can estimate the gaugino mass
of SU(2) as
$$M_2={\sqrt 3 \over f_a}m_{3/2}D_2(T^2,\bar T^2)(b'^2_2-\delta_{GS}^2)
\sin \theta' \cos \theta''.
\eqno(4.4)$$
In the case of the minimal string unification, $\delta^2_{GS}$ is estimated as
$\delta^2_{GS}=0.5\times(b'^2_2-b'^2_3)$ \cite{MSM3}.
Then the values of $T^2=7.5$, 6.5, 5.5 and 4.5 lead to $M_2=63,$ 84, 110
and 140 GeV, respectively, under $\sin \theta'=1/\sqrt 2$ and
$\cos \theta''=1$.
The masses of the singlet superpartners are of the same order as the gaugino
masses.
These spectrum are consistent with ones assumed initially.

It is remarkable that the gaugino mass of SU(3) is different from one of
SU(2) by a factor $(b'^2_3-\delta^2_{GS})/(b'^2_2-\delta^2_{GS})$, which is
equal to --1 in the above example.
The gaugino masses are in general non-universal when
$\sin \theta =0$.
Actually we can obatain large values of $\Delta b'$, e.g.,
$\Delta b'>O(10)$ \cite{MSM2,MSM3} and these values could lead to large
non-universality of the gaugino masses.

We eliminate $\delta_{GS}$ and the T-dependent term of (4.1) using
$\alpha_3$, $\alpha_2$ and $\alpha_1$, so that we have \cite{MSM2}
$$k_1={12\Delta b'{\rm log }(M_{\rm st}^2 / \mu^2)-4B'{\rm log}
 (M_X^2 / M_{\rm st}^2)- 4 \pi\Delta b'\alpha^{-1}_{\rm em}(\mu) \over
 \Delta b'{\rm log }(M_{\rm st}^2 / \mu^2)-4b'^2_2{\rm log}
 (M_X^2 / M_{\rm st}^2)-4 \pi\Delta b'\alpha^{-1}_2(\mu)}-1,
\eqno{(4.5)} $$
where $\Delta b'=b'^2_3-b'^2_2$ and $B'=b'^2_3+b'^2_2=-2$.
Eq.(4.5) is available at $\mu$ where the SUSY is preserved.
We take the example where the masses of the doublet scalar fields are equal
to 3TeV.
We have $\alpha_3^{-1}(3{\rm TeV})=10$, $\alpha_2^{-1}(3{\rm TeV})=31$ and
$\alpha_{\rm em}^{-1}(3{\rm TeV})=125$ in Case III.
In addition to these values, we use $M_X=10^{17.2}$GeV and $\mu=3$TeV so as
to obtain $k_1=1.4$.
It seems reasonable compared with the results of ref.\cite{MSM2,MSM3}.
The minimal string unification with the non-universal soft masses can be
realized for other assignments of the modular weights in the $Z_6$-II
orbifold models.
In the above discussion, we do not take into account the duality anomaly
cancellation condition \cite{Derendinger,Ibanez}, which is used as another
constraint for the realistic models.

The $Z_N \times Z_M$ orbifold models have the rich structure of the modular
weights and the three moduli fields contribute to the threshold corrections.
They can derive the minimal string models under several types of
assignments of the modular weights to the matter fields.


\leftline{\large \bf 5. Conclusions}
\vspace{0.8 cm}

We have studied the soft scalar masses in comparison with the gaugino
masses in the case where the three independent moduli fields as well as
the dilaton field contribute to the SUSY-breaking.
We have showed that the superstring theories can derive the
different order of the non-universality in the scalar partner spectrum.
For such non-universal cases, we have investigated the conditions under
which the modular weights are allowed.
In addition the superstring theories can also obtain the non-universal
gaugino masses.

The non-universality of the soft terms affects phenomenological properties
of the models.
As an example we have studied the gauge coupling unification of the
minimal string models with a certain type of the non-universal soft masses.
We have showed that the minimal string unification with the non-universal
scalar masses is realized in the restricted cases.
It is very important to investigate all the possible models systematically
as refs.\cite{Ibanez,MSM2,MSM3}.
In a similar way, the other cases of the non-universality can be studied.
If we detect the non-universality of the superpartner spectrum in future,
we may constrain promising models in the minimal string models.
It is easy to extend these analysis to the case of extended SUSY models.

Other phenomenological properties are influenced by the non-universality.
For example, the electric dipole moment of the neutron is examined in
ref.\cite{EDMn}.
It is very worthy to study what phenomenological features
are sensitive to the non-universality of the soft terms.
That might lead us to the indirect determination of the superpartner
spectrum.

\vspace{0.8 cm}
\leftline{\large \bf Acknowledgement}
\vspace{0.8 cm}

The authors would like to thank Masahiko~Konmura, Tadao~Suzuki and
Haruhiko~Terao for
useful discussions.
The work of T.K. is partially supported by Soryuushi Shogakukai, and the
work of D.S. is in part supported by a Grant-in-Aid for Scientific
Research from the Ministry of Education, Science and Culture
(\#05640337 and \#06220201).



\newpage
\pagestyle{empty}
\noindent
{\bf \large Table 1. Conditions for real soft scalar masses in $Z_N$
orbifolds }\\
The first column shows the modular weights obtained the $Z_N$ orbifold
models of the second column.
$*$ in the second column represents all the $Z_N$ orbifold
models except $Z_3$ and $Z_7$.
Modular weights forbidden under any condition are indicated by --- in the
third column.

\vspace{5mm}

\begin{tabular}{|c|c|c|}
\hline
Modular Weight & Orbifold & Condition
\\ \hline
(--1,0,0) & & $\sin^2 \theta'\sin^2 \theta'' \leq 1/3$ \\
(0,--1,0) & & $\sin^2 \theta'\cos^2 \theta'' \leq 1/3$ \\
(0,0,--1) & & $\cos^2 \theta' \leq 1/3$ \\ \hline
--(2,2,2)/3 & $Z_3,Z_6$-I& --- \\
--(3,3,2)/4 & $Z_4,Z_8$-I& --- \\
--(2,2,0)/4 & $*$        & $\sin^2 \theta'\leq 2/3$ \\
--(5,5,2)/6 & $Z_6$-I,$Z_{12}$-I & $\sin \theta'=0$ \\
--(5,3,4)/6 & $Z_6$-II  & --- \\
--(4,0,2)/6 & $Z_6$-II  & $\sin^2 \theta'' \leq 1/2$ \\
--(2,0,4)/6 & $Z_6$-II  & $\sin^2 \theta'(2-\sin^2 \theta'')\geq 1$ \\
--(\underline {6,5,4})/7 & $Z_7$     & --- \\
--(\underline {7,3},6)/8 & $Z_8$-I  & --- \\
--(\underline {7,5},4)/8 & $Z_8$-II & --- \\
--(6,2,0)/8 & $Z_8$-II,$Z_{12}$-I & $\sin^2 \theta'(1+2 \sin^2 \theta'')
\leq  4/3$ \\
--(2,6,0)/8 & $Z_8$-II,$Z_{12}$-I & $\sin^2 \theta'(3-2\sin^2 \theta'')
\leq  4/3$ \\
--(\underline {11,5},8)/12 & $Z_{12}$-I & --- \\
--(8,8,8)/12, & $Z_{12}$-I & ---\\
--(\underline {11,7},6)/12 & $Z_{12}$-II & --- \\
--(10,2,0)/12 & $Z_{12}$-II & $\sin^2 \theta'(1+4\sin^2 \theta'') \leq 2$ \\
--(9,9,6)/12 & $Z_{12}$-II & --- \\
--(8,4,0)/12 & $Z_{12}$-II & $\sin^2 \theta'(1+\sin ^2\theta'')\leq 1$ \\
--(4,8,0)/12 & $Z_{12}$-II & $\sin^2 \theta'(2-\sin ^2\theta'')\leq 1$ \\
--(2,10,0)/12 & $Z_{12}$-II & $\sin^2 \theta'(5-4\sin^2 \theta'') \leq 2$ \\
 \hline
\end{tabular}

\newpage
{\bf \large Table 2. Conditions for real soft scalar masses in
$Z_N \times Z_M$ orbifolds }\\
In the second column $*1$ represents $\ob{2}{2},\ob{4}{4},\ob{2}{4},
\ob{2}{6}, \ob{2}{6}',\ob{6}{6}$ and $*2$ represents $\ob{3}{3},\ob{6}{6},
\ob{3}{6}$.
Modular weights forbidden under any condition are indicated by --- in the
third column.
\vspace{5mm}

\begin{tabular}{|c|l|c|} \hline
Modular Weight &\multicolumn{1}{|c|}{Orbifold} & Condition \\ \hline
$-(0,1,1)/2$ & $*1,\ob{3}{6}$
             & $ \sin^2\theta '\sin^2 \theta '' \ge 1/3 $   \\
$-(1,0,1)/2$ & $*1$
             & $\sin^2\theta '\cos^2 \theta '' \ge 1/3 $   \\
$-(1,1,0)/2$ & $*1$
             & $\sin^2\theta ' \le 2/3$                 \\
$-(0,2,1)/3$ & $*2,\ob{2}{6}$
             & $\cos^2\theta '' \le 1/2$   \\
$-(0,1,2)/3$ & $*2,\ob{2}{6}$
             & $\sin^2\theta '(1+\sin^2\theta '') \ge 1$   \\
$-(2,0,1)/3$ & $*2$
             & $\sin^2\theta '' \le 1/2$   \\
$-(2,2,2)/3$ & $*2,\ob{2}{6}'$
             & ---   \\
$-(2,1,0)/3$ & $*2$
             & $\sin^2\theta '(\sin^2\theta ''+1) \le 1$  \\
$-(1,0,2)/3$ & $*2$
             & $\sin^2\theta '(\sin^2\theta ''-2) \le 1$  \\
$-(1,2,0)/3$ & $*2$
             & $\sin^2\theta '(2-\sin^2\theta '') \le 1$  \\
$-(0,3,1)/4$ & $\ob{2}{4},\ob{4}{4}$
             & $\sin^2\theta '(2-\sin^2\theta '') \le 1/3$  \\
$-(0,1,3)/4$ & $\ob{2}{4},\ob{4}{4}$
             & $\sin^2\theta '(2+\sin^2\theta '') \le 5/3$  \\
$-(2,3,3)/4$ & $\ob{2}{2},\ob{4}{4}$
             & ---   \\
$-(5,5,2)/6$ & $\ob{2}{6}',\ob{6}{6}$
             & $\sin^2\theta '=0$  \\
$-(2,5,5)/6$ & $\ob{2}{6}',\ob{3}{6},\ob{6}{6}$
             & $\sin^2\theta '\sin^2\theta''=1$  \\
$-(5,2,5)/6$ & $\ob{2}{6}',\ob{6}{6}$
             & $\sin^2\theta '\cos^2\theta ''=1$  \\
$-(0,5,1)/6$ & $\ob{2}{6},\ob{3}{6},\ob{6}{6}$
             & $\sin^2\theta '(4-5\sin^2\theta '') \le 1$  \\
$-(0,1,5)/6$ & $\ob{2}{6},\ob{3}{6},\ob{6}{6}$
             & $\sin^2\theta '(4+\sin^2\theta '') \ge 3$  \\
$-(3,\underline{5,4})/6$ & $\ob{2}{6},\ob{6}{6}$
             & ---   \\
$-(3,0,1)/4$ & $\ob{4}{4}$
             & $\sin^2\theta '(3\sin^2\theta ''-1) \le 1/3$  \\
$-(3,\underline{3,2})/4$ & $\ob{4}{4}$
             & ---   \\
$-(3,1,0)/4$ & $\ob{4}{4}$
             & $\sin^2\theta '(1+3\sin^2\theta '') \le 4/3$  \\
$-(1,0,3)/4$ & $\ob{4}{4}$
             & $\sin^2\theta '(3-\sin^2\theta '') \ge 5/3$  \\
$-(1,3,0)/4$ & $\ob{4}{4}$
             & $\sin^2\theta '(3-2\sin^2\theta '') \le 4/3$  \\
$-(4,\underline{5,3})/6$ & $\ob{3}{3},\ob{6}{6}$
             & ---   \\
$-(5,1,0)/6$ & $\ob{6}{6}$
             & $\sin^2\theta '(1+4\sin^2\theta '') \le 2$  \\
$-(1,0,5)/6$ & $\ob{6}{6}$
             & $\sin^2\theta '(5-\sin^2\theta '') \ge 3$  \\
$-(1,5,0)/6$ & $\ob{6}{6}$
             & $\sin^2\theta '(5-4\sin^2\theta '') \le 2$  \\ \hline
\end{tabular}

\end{document}